\documentclass[journal]{IEEEtran}

\usepackage{amssymb}
\usepackage{amsmath}
\usepackage{graphicx}
\usepackage{setspace}
\usepackage{color}

\begin{document}
%
\title{Can Visible Light Communications Provide Gb/s Service?}

\author{{Mohammad Noshad,~\IEEEmembership{Student Member,~IEEE}, and Ma\"{\i}t\'{e} Brandt-Pearce,~\IEEEmembership{Senior Member,~IEEE}}
\thanks{Mohammad Noshad (mn2ne@virginia.edu) and Ma\"{\i}t\'{e} Brandt-Pearce (mb-p@virginia.edu) are with Charles L. Brown Department of Electrical and Computer Engineering, University of Virginia Charlottesville, VA 22904.  M. Brandt-Pearce is the formal corresponding author for the paper.}
}

\markboth{}{Shell \MakeLowercase{\textit{et al.}}: Bare Demo of IEEEtran.cls for Journals}

\maketitle

\begin{abstract}
Visible light communications (VLC) that use the infrastructure of the indoor illumination system have been envisioned as a compact, safe, and green alternative to WiFi for the downlink of an indoor wireless mobile communication system.  Although the optical spectrum is typically well-suited to high throughput applications, combining communications with indoor lighting in a commercially viable system imposes severe limitations both in bandwidth and received power. Clever techniques are needed to achieve Gb/s transmission, and to do it in a cost effective manner so as to successfully compete with other high-capacity alternatives for indoor access, such as millimeter-wave radio-frequency (RF). This article presents modulation schemes that have the potential to overcome the many challenges faced by VLC in providing multi Gb/s indoor wireless connectivity.
\end{abstract}


\section{Introduction}
The growth in the use of smartphones and tablets has significantly dominated that of other electronic devices, and this had led to an increasing demand for high-speed data-rates, especially in mobile indoor environments. High-quality applications on these devices will require Gb/s wireless connectivity to the Internet, now only a dream but soon to become  vital. Among the available technologies for providing such high-speed connections, optical wireless communications is a strong contender for next generation indoor interconnection and networking. Visible light is preferred over infrared (IR) communications since it can be integrated with the lighting system and offer a compact, dual-use, energy saving solution. Aside from integrability with the illumination system, visible light communication (VLC) has many  advantages compared with other technologies: radio frequency (RF) interference-free, RF interference immune, safe for human health, and security.

When considering the challenges to commercialization, two key issues are whether the technology can provide a robust solution to the societal need, and whether it can do it better or cheaper than the competition. For high-speed (multi-Gb/p) connectivity, the main current competitor is millimeter-wave RF (e.g., 60 GHz). In this paper we address the first issue by discussing modulation options that address the many technical challenges in implementing VLC at these speeds. We also identify which issues are particular to VLC and which would apply to any approach at these data-rates.

Integrating VLC networks with illumination systems imposes limitations on the types of modulations techniques that can be used. White light emitting diodes (LED) are the most common optical sources proposed for VLC systems, and modulation schemes that can be used with these devices are limited because of the relatively slow rise-time and strong nonlinear behavior. Pulsed techniques are better able to cope with nonlinear effects than subcarrier modulation approaches, yet these do not easily lend themselves to high spectrally efficiency. Dimming is an important feature of indoor lighting systems through which the illumination level can be controlled. Including dimming in VLC system requires further constraints on the modulation schemes that can be used. In this article we discuss various approaches and their ability to address these seemingly contradictory requirements.

Among the different system configurations proposed in \cite{OWC-Configs97}, diffuse link are favored for VLC links since they decrease the link outage caused by shadowing. According to this configuration, the receiver has a large field of view (FOV) so that when the direct path from the source is blocked by an obstacle, the connection can be maintained using the light reflected from the walls. The disadvantage of using a large FOV is two-fold: The received signal can contain  multipath energy with potentially strong magnitude and long delays, and a stronger level of background radiation is collected by the receiver. The power of the background light can be as strong as the received power from the lighting system. The system then becomes both bandwidth and signal-to-noise ratio (SNR) constrained.

In an indoor consumer-centered environment, the downlink from the network to the mobile devices typically demands much higher throughput than the uplink. VLC is used for the downlink only, and another lower data-rate solution is needed for the uplink. Invisible optical bands could be used for the uplink channel in order to avoid self-interference from full-duplex communications. (Using VLC from the mobile device through the LCD display is not technically viable at the moment.) Infrared (IR) and ultraviolet (UV) are two optical candidates for uplink communications. The IR band has higher efficient devices available and a lower impact on human health compared to UV, yet background radiation from the Sun and artificial lights is high in IR. Due to ozone-layer filtering of the Sun, background light is much weaker in the UV 200-280 nm range, and thus UV can provide the same performance as IR with considerably lower transmitted power. Another possibility is to simply rely of low-rate RF for the uplink. In the remainder of the paper, we focus on the downlink because our emphasis is high-throughput;  any technique proposed could equivalently be employed for an IR or UV uplink solution.


Recently a laboratory demonstration of a 1 Gb/s VLC link using multi-input multi-output (MIMO) and orthogonal frequency division multiplexing (OFDM) for an illumination level of 1000 lx and a receiver aperture of 20 cm$^2$ was reported in \cite{OWC-1Gbs-MIMO-OFDM13}. While such examples of high-rate implementations are promising, they do not provide complete solutions. For a commercial VLC system, the receiver should have an aperture size smaller than 0.1 cm$^2$, and the standard illumination level is 400 lx. According to these considerations, the power level received by users in a practical VLC system is on the order of 5 $\mu W$. This illustrates how challenging it is to achieve Gb/s transmission speeds in commercial VLC systems.

The IEEE 802.15.7 standard that was introduced by the Visible Light Communications Consortium (VLCC) proposes one-off keying (OOK), variable pulse-position modulation (VPPM), and color-shift keying (CSK) as modulation techniques for indoor VLC systems. In this standard the highest data-rate envisaged is 96 Mb/s for OOK and CSK, or 24 Mb/s for VPPM. In this article we first discuss challenges to be overcome to increase the data-rate to Gb/s speeds. Then we present newly proposed candidates for modulating LED-based VLC that have the potential to be used for high-speed connection.

\section{Challenges for Modulations in VLC Systems}\label{sec:challenges}
\begin{figure*} [!t]
    \begin{center}
    \scalebox{0.32}{\includegraphics{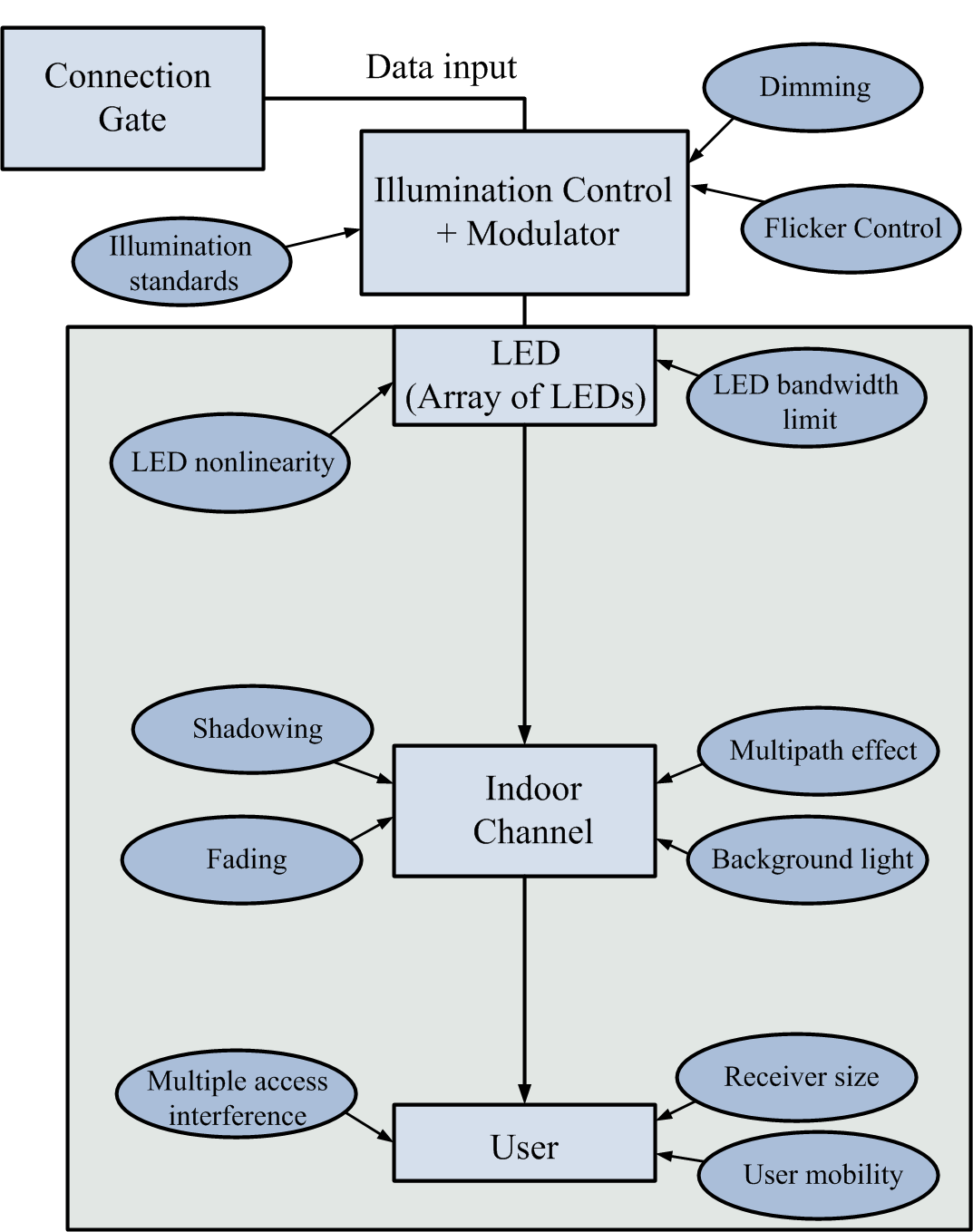}}
    \end{center}
    \vspace*{0.0 in}
    \caption{Configuration of a downlink VLC system using LED arrays, and  challenges affecting each component.}
    \label{VLC}
    \vspace*{-0.0 in}
    \end{figure*}

There are severe limitations on the modulation schemes that can be used in VLC systems, imposed by either illumination features or the devices that are used for simultaneous lighting and communication purposes. Fig.~\ref{VLC} depicts the components of a typical downlink VLC system and their corresponding challenges. An appropriate modulation technique should be able to provide solutions for these issues and fulfill the required constraints. In this section we outline the main design considerations that are required for modulations in VLC.

\subsection{White LEDs}
LEDs are the most likely optical sources for a dual-use lighting and communication application, and thus form the central component of the VLC system transmitter. They are preferred over other lighting sources, such as incandescent and fluorescent sources, since they can respond to faster modulations and support higher data-rates. LEDs are preferred over laser diodes (LD) due to safety regulations, as LDs in the visible range can be harmful to the eyes.
White LEDs are used as arrays to provide the required illuminance in indoor lighting. There are two type of LEDs that use different technologies to generate white light by combining several colors.
In the first technique, an LED emitting blue light is embedded in a layer of yellow phosphor that converts some of the light to longer wavelengths, yellow and red; the result is seen as white light to the human eye. The second type is a trichromatic LED, in which green, blue and red LEDS are integrated into a single device to emit white light. This kind of LED enables easy color rendering by adjusting each color independently. Despite the phosphorescent LEDs having a lower price compared to trichromatic LEDs, the latter are preferred for dual-use since they have a faster rise-time and each color can be modulated independently, tripling the total throughput.

High-speed communications require either high bandwidths (fast rise-times) and/or the use of spectrally-efficient modulation schemes. Commercially viable VLC systems must use LEDs, yet the bandwidth of the existing low-cost devices is limited to a few MHz for phosphor-based LEDs and a few tens of MHz for trichromatic LEDs. In \cite{Indoor-OWC-Obrien-09}, blue-filtering is introduced as an efficient technique for increasing the modulation bandwidth in phosphor-based LED. Also, Le-Minh \cite{VLC-LED-Equlization08} has proposed an equalization technique at the transmitter to considerably increase the net bandwidth of LEDs. Still the source rise-time is insufficient for binary high-speed transmission, requiring the use of higher-order modulation.

As shown in Fig.~\ref{VLC}, the other major limitation imposed by LEDs is their nonlinear transfer function, through which the output power of the LED changes nonlinearly with the modulated input current. This nonlinearity introduces a distortion on the transmitted signal, and this may degrade the performance of the system. Analog or subcarrier modulation schemes suffer from this nonlinear behavior the most. Among the multiple solutions that have been proposed to mitigate this problem, the most efficient  is to limit the modulation index so that high peak amplitudes are avoided, which forces the LED to operate in the linear regime and circumvents the problem. Unfortunately this approach also lowers the effective SNR. Commercially available LED lamps are composed of an array of LEDs, so that they can provide high illumination levels. Below we describe how  to use this structure to avoid some of the nonlinear distortion of single LED transmitters in multi-amplitude modulation schemes.

Note that transmitter rise-time and amplifier nonlinearity are likely to be encountered by any technology attempting high throughput communications. However, the limits seen by white LEDs are particularly severe, making the design of high-speed modulation schemes especially challenging.

\subsection{Features of the Lighting System}

One of the attractive aspects of VLC is the possibility to design dual-use systems satisfying both indoor illumination and communications needs. While this approach saves energy, requirements of the lighting system impose extra requirements on the communication system.

Dimming is an important feature of indoor lighting systems through which the illumination level can be controlled by the user. To support dimming, a practical VLC system should be able to operate at various optical peak to average power ratios (PAPR) so that, for a fixed peak power LED, the average power, which is proportional to the illumination, can be regulated. Continuous current reduction (CCR) and pulse-width modulation (PWM) are two techniques that have been proposed for dimming in indoor VLC systems \cite{OWC-Dimming-OFDM12}; these techniques require large bandwidths, and are therefore not suitable for high-rate systems. Dimming to low light-levels also reduces the power received at the mobile, potentially affecting the data-rate that is achievable. Dimmable VLC systems must be adaptive, adding complexity to the entire system.

Flicker is a fluctuation of the illumination that can be perceived by human eyes and must be avoided. After long-term exposure, flicker can be harmful to the eye and affect eyesight. In VLC systems, since the lighting is integrated with the communication system, an inappropriate modulation scheme can cause variations in the average transmitted power, and impose fluctuations on the brightness of the LEDs. Therefore, constraints need to be applied to modulation techniques that are aimed at dual-use VLC systems. The flicker becomes important when the data-rate is low, or the lights are dimmed to a low illumination level. Even though the IEEE 802.15.7 standard has devised some techniques to alleviate flicker in VLC, suitable modulation schemes are still required to have a constant average power over several symbols.



\subsection{Indoor VLC Channel}

The channel impulse response of a VLC system is composed of two parts: a line-of-sight (LOS) part that is the response received from the direct path to the closest light source, and a non-line-of-sight (NLOS) part that is the response received after reflection from walls and other objects. The latter can also be considered as  multipath. The LOS part of the impulse response is usually short and sharp, and the NLOS part is instead broad, depending of the reflectivity of the walls and the size of the room. This broad response causes an intersymbol interference (ISI) effect on a high-speed transmitted data stream and degrades the quality of the received signal. Therefore, modulations that are less susceptible to ISI are of interest in VLC.

    \begin{figure*} [!t]
    \begin{center}
    \scalebox{0.21}{\includegraphics{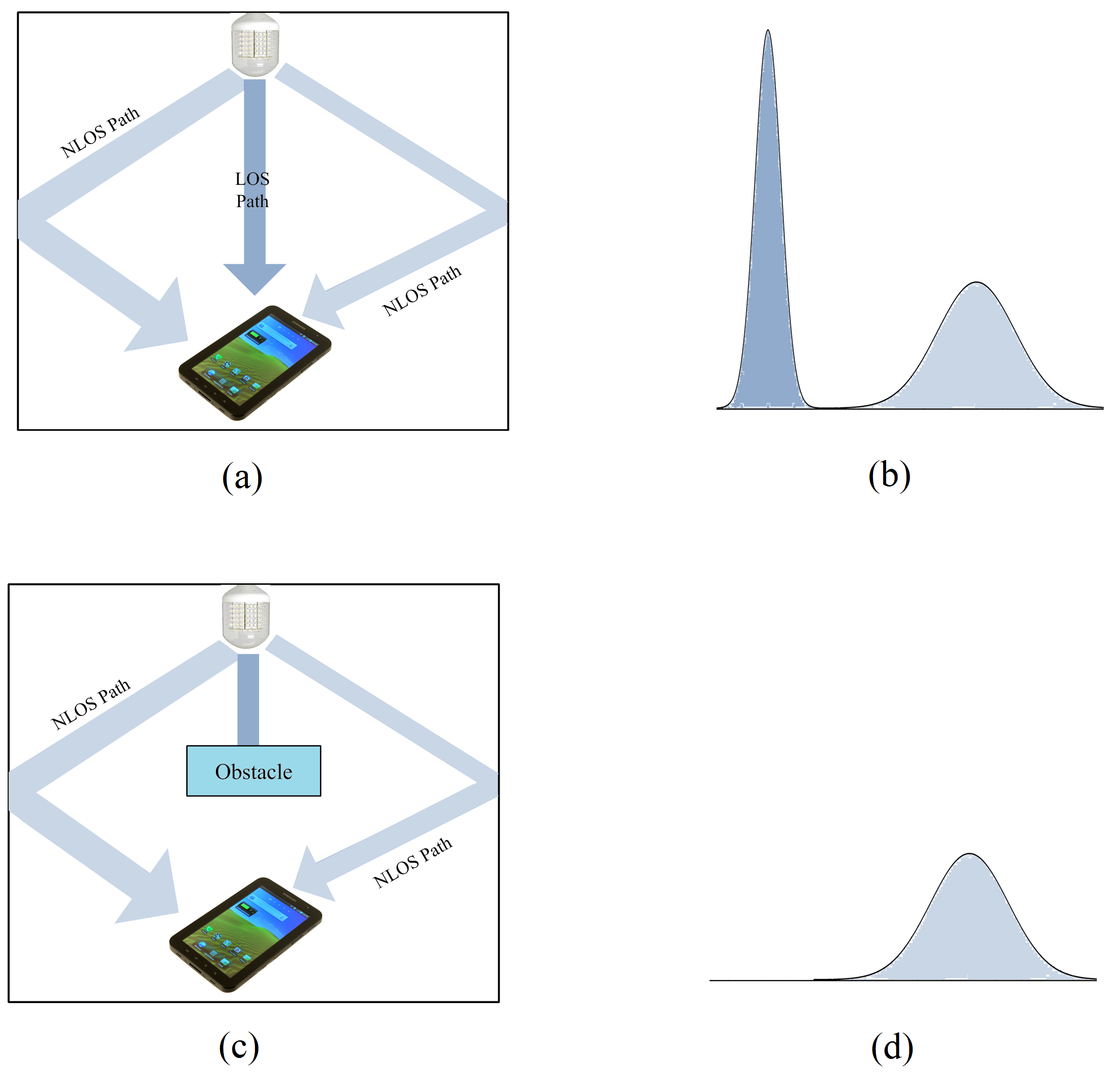}}
    \end{center}
    \vspace*{0.0 in}
    \caption{(a) A VLC system that receives data from both LOS and NLOS paths, and (b) its corresponding impulse response. (c) Shadowing effect in VLC when the direct path is blocked, and (d) the impulse response in the shadowing case.}
    \label{LOS-NLOS}
    \vspace*{-0.0 in}
    \end{figure*}

Another important channel effect in VLC is shadowing, which is a blocking of the direct path from the lighting source to the photodetector. In VLC systems, the LOS part has a high blocking probability because visible light radiation does not penetrate opaque objects, unlike RF waves. Shadowing can be caused by obstructions in the indoor environment or by people moving, making the channel time-varying. As is shown in Fig.~\ref{LOS-NLOS}, in the shadowed situations the impulse response of the channel has only the NLOS part, and data is retrieved using this part. Since the VLC channel response can change very fast, modulation schemes that do not require a threshold to make decision are preferred.

Both of these channel effects are present to some extent in RF communications. The added problem of fading due to multipath-wave cancelation so prevalent in RF is not experienced in VLC since the optical sources are intensity modulated (IM) and incoherent.

The last problem encountered in VLC is potentially strong background light, especially if direct sunlight is in the FOV of the photodetector. Illumination levels are strong enough that in the absence of background light the SNR is quite high. Unlike RF where interference can be mitigated by using MIMO processing, background light increases the shot noise of the system irreparably.


\section{Modulation Schemes}\label{sec:modulations}

More sophisticated modulation schemes than proposed in the current standard are needed to allow multi-Gb/s VLC systems to simultaneously satisfy the requirements of the illumination system, and device and channel constraints mentioned above. As the symbol-rate is limited by both source and channel properties, increasing the throughput requires increasing the modulation constellation size. In IM point-to-point systems, the only degree-of-freedom is the instantaneous power, and thus either subcarrier-based or multi-amplitude modulations are possible. Spectrally efficient modulation candidates are discussed below.

\subsection{OFDM}
OFDM is a popular multi-carrier modulation technique developed for RF systems able to significantly increase the data-rate in bandwidth-constrained channels. Because of its outstanding performance in RF systems, modified forms of it, such as DC biased optical OFDM (DCO-OFDM) and asymmetrically clipped optical OFDM (ACO-OCDMA), have been proposed for use in indoor VLC systems. Bit-rates of 1 Gb/s in combination to MIMO have been demonstrated.

Several challenges remain in using OFDM for VLC. The tails of the impulse response can be quire long, requiring a long cyclic prefix for OFDM to work, and this reduces the throughput. Dimming is another challenge in utilizing OFDM in indoor VLC. OFDM has a naturally high PAPR, but it is not easily controllable. A solution for  embedding the dimming function in OFDM is to combine it with PWM, as suggested in \cite{OWC-Dimming-DMT11}, but this approach limits the data-rate and can cause flicker.

By far the worst problem for OFDM is the LED nonlinearity since, due to the large PAPR, it can severely distort the output signal. As alternatives to reducing the modulation index discussed above, various other compensation techniques have been proposed to mitigate nonlinearity-induced distortions \cite{OWC-Nonlinearity-OFDM12}. While some techniques try to modify  OFDM to make it resistant against LED nonlinearity, others introduce compensators to linearize the overall transfer-function.
One possible approach that has yet to be explored it to use the LED array elements separately to transmit the OFDM signal. An $N$-element array can transmit an $M$-subchannel OFDM signal by assigning $N/M$ LEDs to each subcarrier, thereby reducing the PAPR to each LED, and eliminating the nonlinear degradation.


\subsection{Spatial Modulation}

To side-step the single-degree-of-freedom limitation of IM, a multi-point system can be used. The resulting so-called spatial modulation (SM) increases the data transmission rate by using multiple physically-separated transmitters and multiple  physically-separated receivers. SM can either be used as a stand-alone modulation or be considered as a form of multiple-input multiple-output (MIMO) that can be combined with other modulation schemes to increase the bit-rate. A demonstration of the integration of SM with OFDM was presented in \cite{VLC-SM-OFDM12}. $M$-ary SM is attractive as it increases the spectral efficiency without requiring additional bandwidth; it also increases the total received power, and it does not rely on a threshold. In combination with OFDM, it can provide data rates up to 1 Gb/s. SM is also helpful in supporting simultaneous access in multiuser VLC networks.

The main drawback of SM in VLC system is its susceptibility to shadowing and multipath interference. In a highly dispersive indoor channel, the signals received via different paths can cause severe interference between transmissions at the receivers. Note that while rich scattering channels resulting from NLOS conditions can help RF systems by providing diversity, in IM optical systems they primarily cause interference. An effective technique to avoid this problem may be to narrow the receivers' FOV, which would reduce interference at the expense of increasing the vulnerability to shadowing, i.e., increasing the blocking probability of the receiver. Furthermore, for commercially viable built-in VLC transceivers, the size of the devices needs to be fairly small, and a SM receiver requiring multiple detectors may not be practical.

\subsection{Pulse position based modulation}

    \begin{figure} [!t]
    \begin{center}
    \scalebox{0.27}{\includegraphics{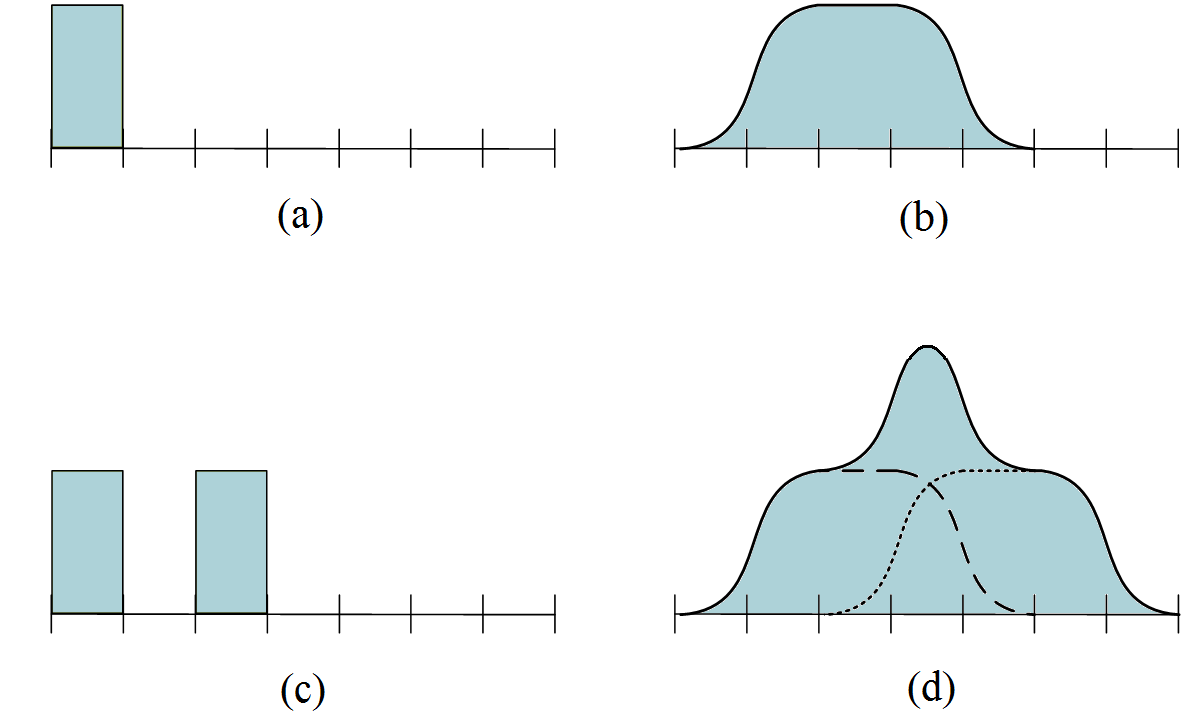}}
    \end{center}
    \vspace*{0.0 in}
    \caption{A illustration of applying overlapping pulses technique to PPM and a multipulse modified PPM.}
    \label{Overlapping}
    \vspace*{-0.0 in}
    \end{figure}

Pulse-position based modulation techniques are traditionally the most popular option for modulating LEDs. These modulation schemes can be easily implemented by turning on and off the LED at predetermined time intervals. They are therefore not affected by the LED nonlinearity, but suffer from the slow rise-time.

To reduce the limitation imposed by the small white LED bandwidths, pulse-position based modulations can use wider pulses than their time-slots would allow, i.e., the so-called overlapping-pulse technique, in order to provide higher-speed transmission of information. The LEDs are effectively modulated with pulses shorter than their response time. Fig.~\ref{Overlapping}(a)(b) illustrates this technique. For pulse-position based modulations that have more than one pulse per symbol time, the same technique can also be used, as shown in Fig.~\ref{Overlapping}(c)(d). In this case the output light can become multi-level. To avoid incurring nonlinear degradation, we can again use the LED array structure: separate LEDs can be used to generate the various pulses at different times. This technique can be applied to any pulse-position based modulation.

    \begin{table*} [!t]
    \caption{Pulse position based schemes with $Q$ time slots and $N$ levels}
    \begin{center}
    \scalebox{0.27}{\includegraphics{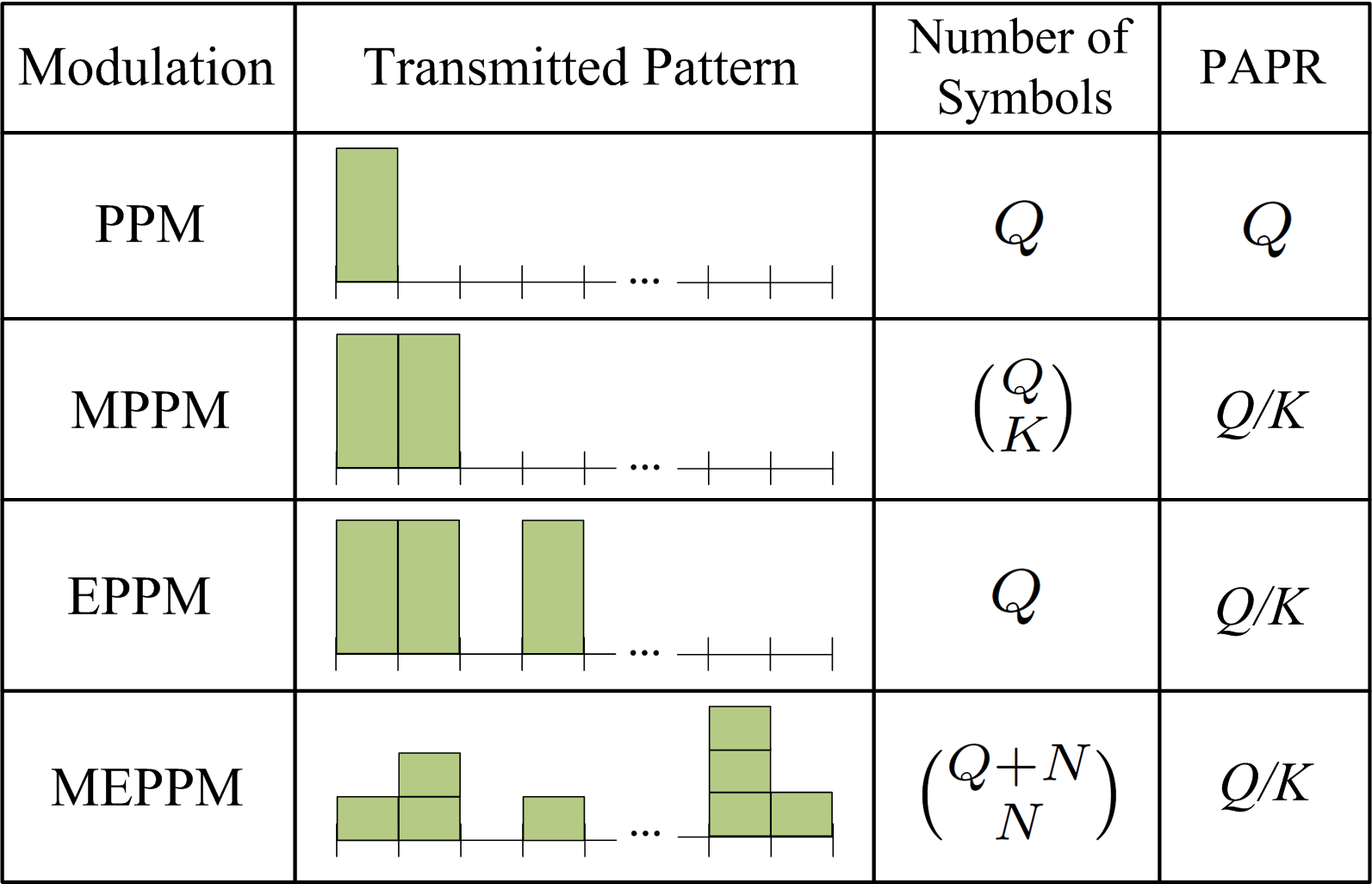}}
    \end{center}
    \label{PPM-Schemes}
    \end{table*}
    \begin{table*} [!t]
    \caption{Comparison between different modulation schemes}
    \begin{center}
    {\includegraphics[width=6.5in]{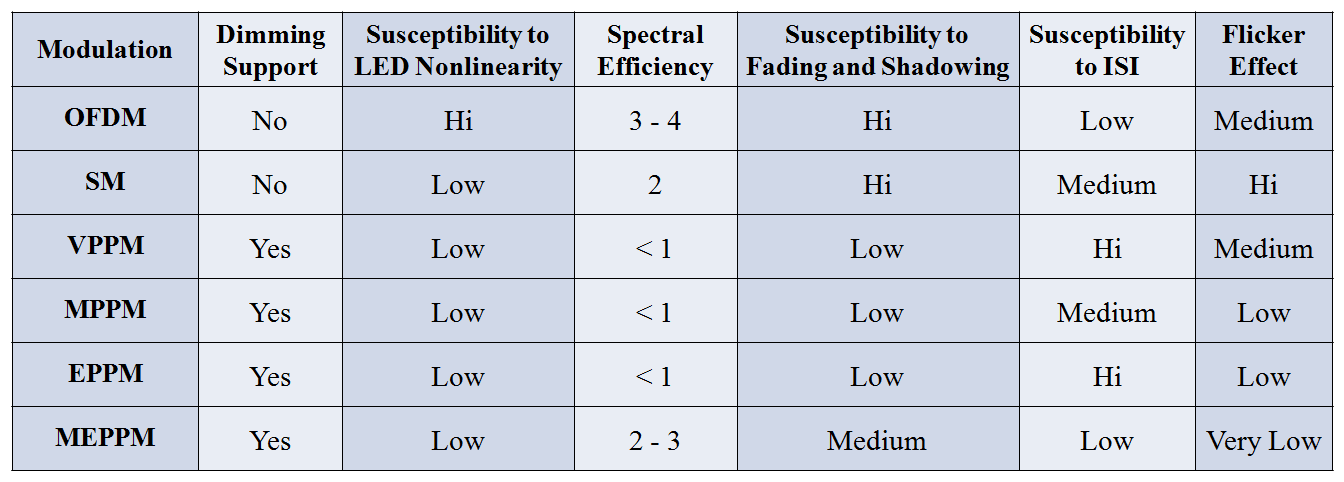}}
    \end{center}
    \label{Modulations}
    \end{table*}

\subsubsection{PPM}

Standard pulse-position modulation (PPM) is a conventional modulation scheme in which the symbol duration is divided into $Q$ equal time-slots and a single pulse is transmitted in one of these time-slots. Hence, the symbols are identified by the position of the pulse. The simple implementation of PPM and its threshold-free detection have made it a popular technique for optical communications. But as the constellation size increases the spectral efficiency decreases, making PPM inappropriate for high-rate transmission. PPM can used to combat strong background light if the system is average-power limited. In peak-power limited systems, increasing $Q$ only reduces the received energy per symbol. Modified and generalized forms of PPM are needed to address these two problems.

\subsubsection{MPPM}

Multipulse PPM (MPPM) has been proposed to increase the spectral-efficiency of PPM by transmitting multiple pulses in each symbol-time. In this scheme, every sequence of length $Q$ containing $K$ pulses is considered as a symbol. This modulation technique has the potential to achieve spectral efficiencies close to one. The constellation of MPPM symbols has a minimum Hamming distance of 2, and hence the performance improvement over PMM is due only to the increase in the constellation size, not the distance between symbols. A second difficulty in MPPM is finding a good bit-to-symbol mapping; a random mapping is usually implemented to avoid complexity.

\subsubsection{EPPM}
Expurgated PPM is a modified form of PPM proposed to enhance its performance in optical systems where sources are peak-power limited \cite{EPPM12}. In this scheme, symbols of MPPM are expurgated in order to maximize the Hamming distance between symbols. The structure of EPPM is as simple as PPM since the symbols are cyclic shifts of each other and the encoder and decoder can be implemented using shift-registers. EPPM can be used in VLC requiring a dimming feature by changing the length of the symbols ($Q$) and the number of pulses per symbol ($K$), setting the PAPR to any desired value while maintaining the large minimum distance property. Yet EPPM still suffers from the same low spectral-efficiency as PPM.

In a dispersive channel the impact of ISI on EPPM is similar to PPM: the multipath components of the received signal correspond to a wrong symbol. Applying an interleaver at the transmitter can considerably mitigate this interference effect on EPPM and decrease the error probability, an option not available to PPM.
An additional advantage of EPPM over PPM is its potential to mitigate flicker since it transmits multiple pulses over a symbol period instead of one. By increasing the number of time-slots in each symbol, their duration becomes smaller, and the light intensity fluctuations become less discernible to the human eye.

\subsubsection{Multilevel EPPM}

The above modulation schemes all have spectral-efficiency less than unity because they are two-level. As mentioned earlier, increasing the spectral efficiency requires exploiting the one dimension available to IM systems, the intensity. The simplest multi-level modulation is pulse amplitude modulation (PAM), which is inappropriate for potentially-shadowed wireless systems since it is very sensitive to threshold levels. A better option is to use a multilevel forms of EPPM, as introduced in \cite{Multilevel-EPPM12}, in order to increase the constellation size and hereby provide higher data-rates. Symbols of MEPPM are constructed as  linear combinations of $N$ EPPM symbols (or their complements). Similar to EPPM, multilevel EPPM (MEPPM) is able to support a wide range of PAPRs, and can transmit high speed data  even in highly dimmed scenarios. The same interleaving technique described for EPPM can also be used for MEPPM to decrease the multipath effect in dispersive VLC channels. Using LED array lamps gives us the opportunity to employ multi-level modulation techniques but still on-off modulating each individual device. Therefore the technique is impervious to the LED nonlinearity.

From the flicker perspective, MEPPM is even better than EPPM in mitigating the flicker effect since the presence of multiple light levels per symbol makes the light intensity change on average less severe from one time-slot to another. By increasing the number of levels and decreasing the duration of time slots the illumination variations become imperceptible to the eye.

Table.~\ref{PPM-Schemes} shows the constellation size and PAPR for the various pulse-position based techniques. EPPM and MEPPM are the only schemes in which the PAPR can be chosen independent from the constellation size. For MEPPM, the spectral efficiency can be larger than unity, making it most applicable to high data-rate VLC transmission.

\subsection{Comparison Between Modulation Schemes}

A comparison between the modulation schemes is given in Table.~\ref{Modulations}. According to this table, MEPPM and OFDM are the candidates that have the highest potential to provide a Gb/s transmission in VLC links. OFDM can certainly achieve a higher spectral efficiency in the absence of any source and channel impairments. Because of MEPPM's insensitivity to nonlinearity and its ability  to support a wide range of dimming levels, it seems to be better suited to indoor VLC. By using the overlapping-pulse  and pulse-interleaving techniques, MEPPM may be able to support even higher data-rates in source and channel band-limited systems.

To help answer the question posed in the article's title of whether Gb/s throughputs are possible using VLC, let us consider some real numbers. To obtain a throughput of 1~Gb/s using a tricolor LED, suppose we require an average data-rate of 333 Mb/s per color. Using a pulse-overlapping factor of 10, MEPPM would have to provide a spectral efficiency of 3 bits/second/Hz for currently available devices. A system using $Q=7$ and $N=21$ would be able to satisfy the source bandwidth requirements. Assuming a 400 lx level of illumination, this modulation can achieve a BER of $3\times10^{-3}$. Using SM together with MEPPM, this data-rate can be increased to a few Gb/s.

\section{Conclusion}

In this article we address the question of whether VLC can provide high data rate transmission in indoor environments in a practical and cost-effective manner for commercialization. Challenges that must be addressed in order for a modulation scheme to be used in VLC are discussed. Multiple modulation techniques are explored that are suitable for VLC, two of which, OFDM and MEPPM, are found to provide the spectral efficiency required for Gb/s throughputs.

\section{Acknowledgment}
This research was funded in part by the National Science Foundation (NSF) under grant number ECCS-0901682.

\bibliographystyle{IEEEtran}
\bibliography{EPPM}

\begin{thebibliography}{10}
\providecommand{\url}[1]{#1}
\def\UrlFont{\rmfamily}
\providecommand{\newblock}{\relax}
\providecommand{\bibinfo}[2]{#2}
\providecommand\BIBentrySTDinterwordspacing{\spaceskip=0pt\relax}
\providecommand\BIBentryALTinterwordstretchfactor{4}
\providecommand\BIBentryALTinterwordspacing{\spaceskip=\fontdimen2\font plus
\BIBentryALTinterwordstretchfactor\fontdimen3\font minus
  \fontdimen4\font\relax}
\providecommand\BIBforeignlanguage[2]{{%
\expandafter\ifx\csname l@#1\endcsname\relax
\typeout{** WARNING: IEEEtran.bst: No hyphenation pattern has been}%
\typeout{** loaded for the language `#1'. Using the pattern for}%
\typeout{** the default language instead.}%
\else
\language=\csname l@#1\endcsname
\fi
#2}}

\bibitem{OWC-Configs97}
J.~Kahn and J.~Barry, ``Wireless infrared communications,'' \emph{Proc. IEEE},
  vol.~85, no.~2, p. 265–298, Feb 1997.

\bibitem{OWC-1Gbs-MIMO-OFDM13}
A.~Azhar, T.~Tran, and D.~O'Brien, ``A {Gigabit/s} indoor wireless transmission
  using {MIMO-OFDM} visible-light communications,'' \emph{IEEE Photon. Tech.
  Lett.}, vol.~25, no.~2, pp. 171--174, 2013.

\bibitem{Indoor-OWC-Obrien-09}
L.~Zeng, D.~O'Brien, H.~Minh, G.~Faulkner, K.~Lee, D.~Jung, Y.~Oh, and E.~T.
  Won, ``High data rate multiple input multiple output ({MIMO}) optical
  wireless communications using white {LED} lighting,'' \emph{IEEE J. Select.
  Areas Commun.}, vol.~27, no.~9, pp. 1654 -- 1662, 2009.

\bibitem{VLC-LED-Equlization08}
H.~L. Minh, D.~O'Brien, G.~Faulkner, L.~Zeng, K.~Lee, D.~Jung, and Y.~Oh, ``80
  {Mbit/s} visible light communications using pre-equalized white {LED},''
  \emph{34th European Conference on Optical Communication (ECOC)}, 2008.

\bibitem{OWC-Dimming-OFDM12}
I.~Stefan, H.~Elgala, and H.~Haas, ``Study of dimming and {LED} nonlinearity
  for {ACO-OFDM} based {VLC} systems,'' \emph{Proceeding of IEEE Wireless
  Communications and Networking Conference}, pp. 990--994, Apr. 2012.

\bibitem{OWC-Dimming-DMT11}
G.~Ntogari, T.~Kamalakis, J.~Walewski, and T.~Sphicopoulos, ``Combining
  illumination dimming based on pulse-width modulation with visible-light
  communications based on discrete multitone,'' \emph{IEEE J. Opt. Commun.
  Netw.}, vol.~3, no.~1, pp. 56--65, 2011.

\bibitem{OWC-Nonlinearity-OFDM12}
R.~Mesleh, H.~Elgala, and H.~Haas, ``{LED} nonlinearity mitigation techniques
  in optical wireless {OFDM} communication systems,'' \emph{IEEE J. Opt.
  Commun. Netw.}, vol.~4, no.~11, pp. 865--875, Nov. 2012.

\bibitem{VLC-SM-OFDM12}
X.~Zhang, S.~Dimitrov, S.~Sinanovic, and H.~Haas, ``Optimal power allocation in
  spatial modulation {OFDM} for visible light communications,'' \emph{75th IEEE
  Vehicular Technology Conference (VTC Spring)}, 2012.

\bibitem{EPPM12}
M.~Noshad and M.~Brandt-Pearce, ``Expurgated {PPM} using balanced incomplete
  block designs,'' \emph{IEEE Commun. Lett.}, vol.~16, no.~7, pp. 968--971,
  2012.

\bibitem{Multilevel-EPPM12}
------, ``Multilevel pulse-position modulation based on balanced incomplete
  block designs,'' \emph{Proceding of IEEE Global communications conference
  ({GLOBECOM})}, Anaheim, CA, Dec. 2012.

\end{thebibliography}

\begin{biography}[{\includegraphics[width=1in,height=1.25in,clip,keepaspectratio]{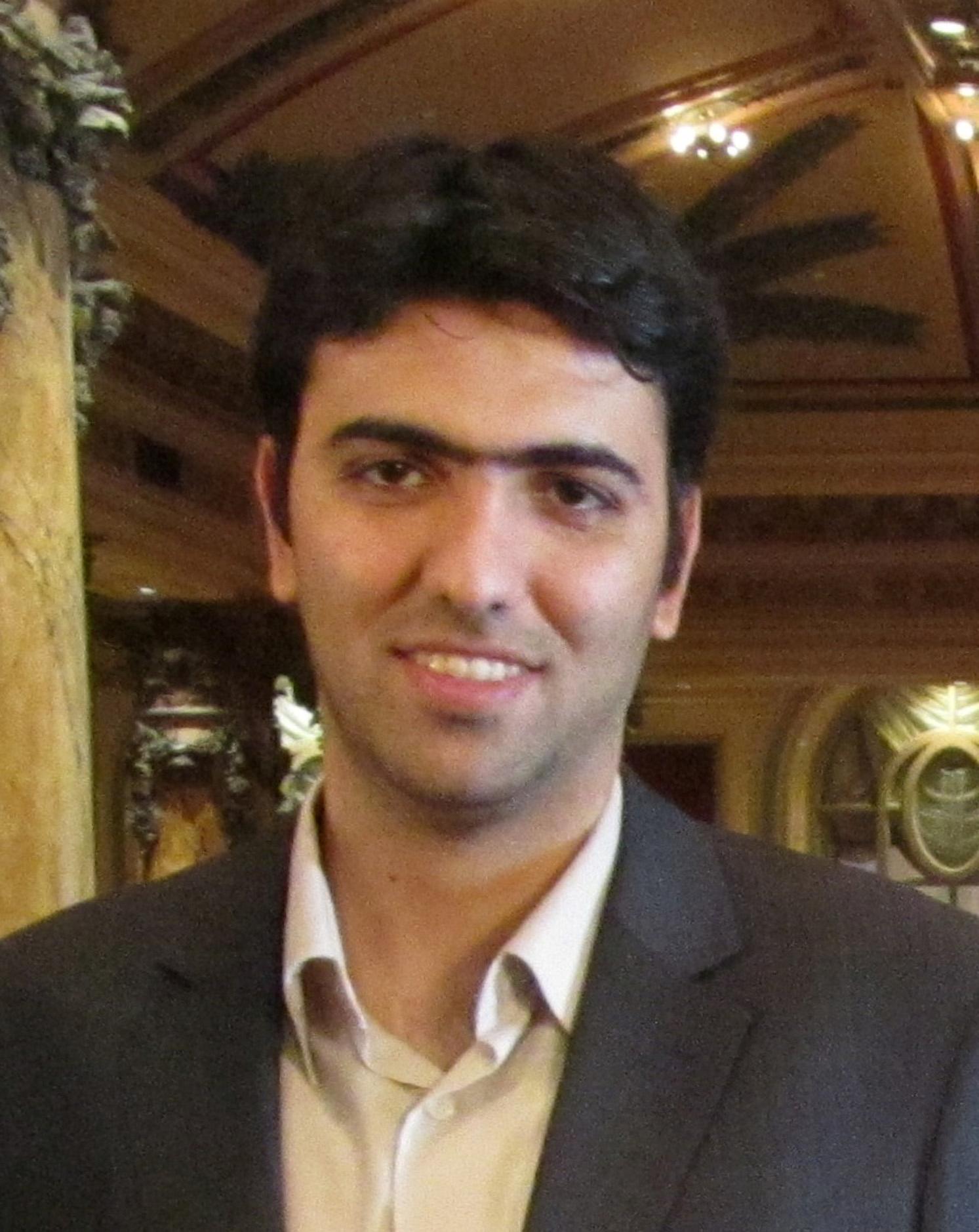}}]{Mohammad Noshad}
(S'07) received the B.Sc. degree from University of Tabriz, Tabriz, Iran, in 2007, and the M.Sc. degree from Sharif University of Technology (SUT), Tehran, Iran, both in electrical engineering. Since January 2011, he has been working towards the Ph.D. degree in the Charles L. Brown Department of Electrical and Computer Engineering at University of Virginia, Charlottesville, VA. From May 2010 to December 2010 he was a researcher in i2cat foundation in Barcelona, Spain.

M. Noshad received ``Louis T. Rader Graduate Research Award" from Electrical and Computer Engineering Department, University of Virginia, in 2012. He is also the recipient of the 2012 ``Charles L. Brown Fellowship for Excellence", as well as the ``TRANE Graduate Fellowship" and ``University of Virginia Engineering Foundation Fellowship". He received the ``Best Paper Award" at the IEEE Globecome 2012. His research interests include coding and modulation, free-space optical communications, network coding and combinatorial designs.
\end{biography}

\begin{biography}[{\includegraphics[width=1in,height=1.25in,clip,keepaspectratio]{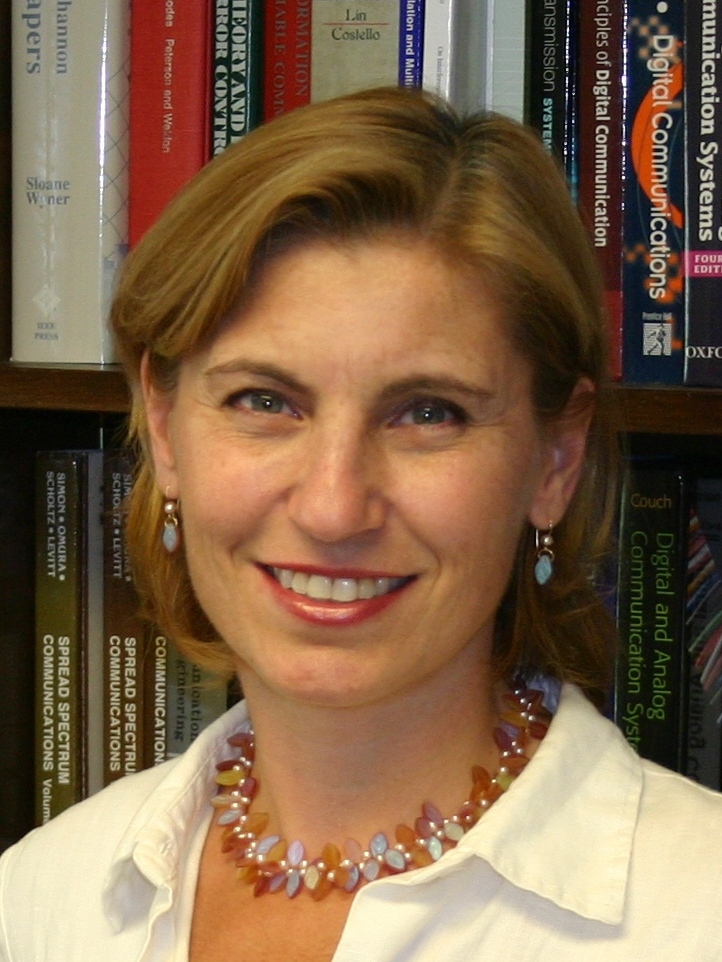}}]{Ma\"{\i}t\'{e} Brandt-Pearce}
(SM'99) received her B.S. in Electrical Engineering from Rice University in 1985. She completed an M.E.E. in 1989 and a Ph.D. in Electrical Engineering in 1993, both also from Rice University. Dr. Brandt-Pearce is currently a professor in the Charles L. Brown Department of Electrical and Computer Engineering at the University of Virginia. Her research interests lie in the mathematical and numerical description and optimization of systems with multiple simultaneous components from different sources and corrupted by non-Gaussian noise. This interest has found applications in a variety of research projects including spread-spectrum multiple-access schemes, multiuser demodulation and detection, study of nonlinear effects on fiber-optic multiuser/multichannel communications, optical networks subject to physical layer degradations, free-space optical multiuser communications, biomedical data processing, and radar signal processing and tracking of multiple targets. Dr. Brandt-Pearce has over a hundred major journal and conference publications.
\end{biography}

\end{document}